\documentclass[aps,PRB,twocolumn,superscriptaddress,preprintnumbers]{revtex4-2}
\usepackage[T1]{fontenc}
\usepackage{times}
\usepackage{graphicx,color}
\usepackage{amsfonts,amsmath,amssymb,amsbsy}
\usepackage[colorlinks=true, linkcolor=blue, citecolor=blue, urlcolor=blue]{hyperref}
\usepackage{float}

\begin{document}

\title{Cavity-induced topological edge and corner states}
\author{Motohiko Ezawa}
\affiliation{Department of Applied Physics, University of Tokyo, Hongo 7-3-1, 113-8656,
Japan}

\begin{abstract}
We investigate a two-level system with alternating XX coupling in a photon
cavity. It is mapped to a free boson model equally coupled to a photon,
whose interaction is highly nonlocal. Some intriguing topological phenomena
emerge as a function of the photon coupling. The photon energy level
anticrosses the zero-energy topological edges at a certain photon coupling,
around which the symmetric edge state acquires nonzero energy due to the
mixing with the photon. Furthermore, the photon state is transformed into
the topological zero-energy edge or corner state when the photon coupling is
strong enough. It is a cavity-induced topological edge or corner state. On
the other hand, the other topological edge or corner states do not couple
with the photon and remains at zero energy even in the presence of the
cavity. We analyze a cavity-induced topological edge state in the
Su-Schrieffer-Heeger model and a cavity-induced topological corner state in
the breathing Kagome model.
\end{abstract}

\maketitle

\textbf{Introduction: } Cavity quantum-electrodynamics (QED)\ is a field of
studying the coupling effect between atoms with discrete levels and a photon%
\cite{Pelli,Ima,Ye,Zheng,BlaisA}. For example, by placing two mirrors in
parallel, the energy of photon is quantized and cavity QED is realized.
Cavity QED is also realized in superconducting qubits based on Josephson
junctions\cite{Gu,Blais}. The coupling constant between the discrete levels
and a photon is largely enhanced experimentally\cite%
{Boura,Niem,Kock,Diaz,AshidaL21}.

Recently, cavity quantum materials attract much attention\cite{CavMatRev},
where a material instead of an atom is deposited in the mirrors.
Superconductivity\cite{Schla}, ferroelectricity\cite{Ashida}, photon-magnon
coupling\cite{Soykal} and quantum Hall effects\cite{Kivis,Scalari,XWang} are
studied in the context of cavity quantum materials. A fermionic
Su-Schrieffer-Heeger (SSH) model coupled with a photon is studied\cite%
{Dmy,Perez,WWang,Shaf}. Topological properties of interacting atoms in the
cavity are also discussed\cite{Down,Nie1,Nie2,CYWang,Saug}, where
superconducting qubits enable us to realize strong inter-atomic couplings%
\cite{Nie1,Nie2}.

In this paper, we investigate the topological zero-energy edge and corner
states in cavity-coupled arrays of an interacting two-level system. It is
mapped to a hopping model with an additional site representing a photon,
which coupled to all of the other sites equally. We analyze a dimerized
one-dimensional chain and a breathing Kagome model with alternating XX
interactions. We show that they are mapped to the SSH model and the
breathing Kagome second-order topological insulator model coupled with an
additional photon site, respectively.\ Intriguing phenomena occur on the
photon state and a topological edge or corner state. As the coupling with
the cavity increases, the photon state is\ smoothly transformed into the
symmetric edge state or the C$_{\text{3}}$-symmetric corner state, while the
latter edge state or corner state is smoothly transformed into the photon
state. We call it the cavity-induced edge or corner state. On the other
hand, the other topological edge state and corner states remain as they are
because they do not couple with the photon.

\begin{figure}[t]
\centerline{\includegraphics[width=0.48\textwidth]{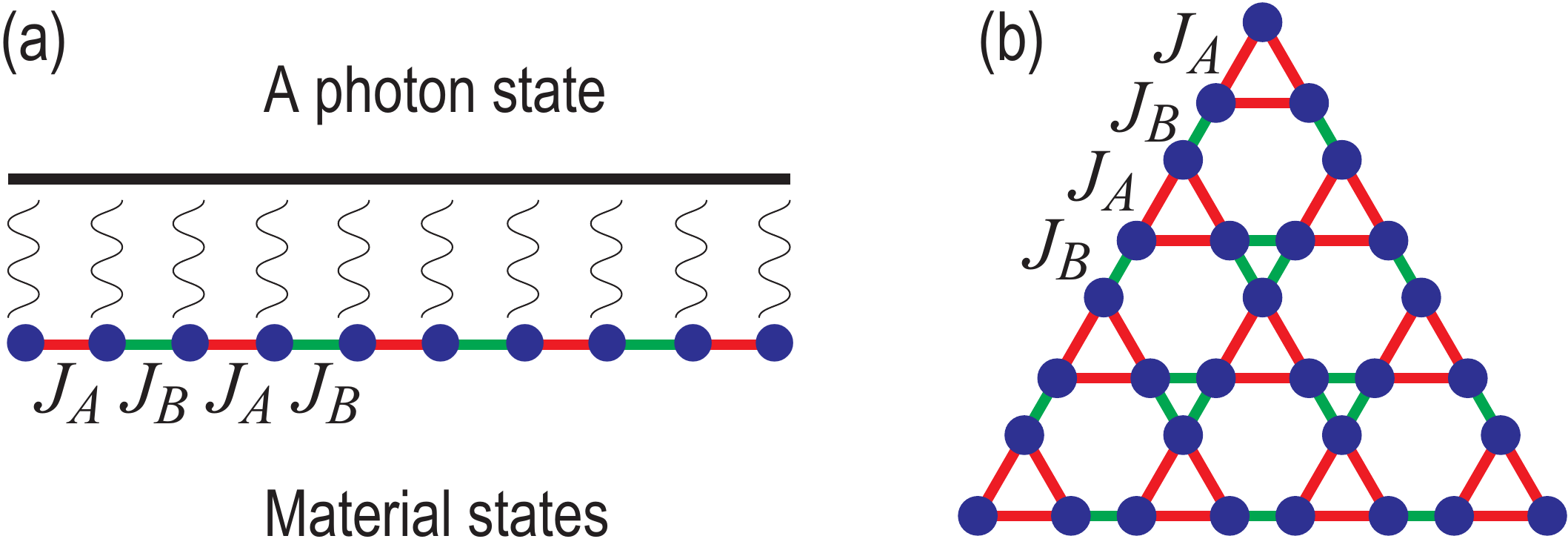}}
\caption{(a) Illustration of the SSH model with each site equally coupled
with a single photon. (b) That of the breathing Kagome lattice.}
\label{FigKagomeNanodiskIllust}
\end{figure}

\textbf{Cavity pseudospin system: } Cavity QED with the XX interaction is
described by%
\begin{equation}
H=\hbar \omega _{0}\hat{a}^{\dagger }\hat{a}+H_{\text{spin}}+\hbar
g\sum_{\alpha }\left( \hat{a}^{\dagger }\sigma _{\alpha }^{-}+\hat{a}\sigma
_{\alpha }^{+}\right) ,  \label{HJS}
\end{equation}%
together with the pseudospin Hamiltonian%
\begin{equation}
H_{\text{spin}}=\hbar \omega _{\text{s}}\left( \sum_{\alpha }\frac{\sigma
_{\alpha }^{+}\sigma _{\alpha }^{-}}{2}-1\right) -\sum_{\alpha ,\beta
}J_{\alpha \beta }\left( \sigma _{\alpha }^{-}\sigma _{\beta }^{+}+\sigma
_{\alpha }^{+}\sigma _{\beta }^{-}\right) ,  \label{HMag}
\end{equation}%
where $\alpha $ and $\beta $ are the indices of atoms, $\sigma $ is a
pseudospin operator describing the two levels of the atom, and $J_{\alpha
\beta }$ is the exchange coupling constant. We take $0<\omega _{0}<\omega _{%
\text{s}}$ and have subtracted the energy of the single spin $\hbar \omega _{%
\text{s}}$ in (\ref{HMag}). The Hamiltonian (\ref{HJS}) is called the
Jaynes-Cummings model\cite{JC} without the exchange interaction terms ($%
J_{\alpha \beta }=0$). The exchange interaction terms are realized based on
superconducting qubits\cite{Nie1,Nie2}.

There is a conserved quantity $\left[ H,N\right] =0$ with the total number $%
N=\hat{a}^{\dagger }\hat{a}+\sum_{\alpha }\sigma _{\alpha }^{+}\sigma
_{\alpha }^{-}$. We consider the case $N=1$. In this case, the system (\ref%
{HJS}) together with (\ref{HMag}) is equivalently described by the bosonic
Hamiltonian%
\begin{eqnarray}
H_{\text{boson}} &=&-\sum_{\alpha ,\beta }J_{\alpha \beta }\left( \hat{b}%
_{\alpha }^{\dagger }\hat{b}_{\beta }+\hat{b}_{\beta }^{\dagger }\hat{b}%
_{\alpha }\right) +\sum_{\alpha }\hbar \omega _{\text{s}}\hat{b}_{\alpha
}^{\dagger }\hat{b}_{\alpha }  \notag \\
&&+\hbar g\sum_{\alpha }\left( \hat{a}^{\dagger }\hat{b}_{\alpha }+\hat{b}%
_{\alpha }^{\dagger }\hat{a}\right) +\hbar \omega _{0}\hat{a}^{\dagger }\hat{%
a}-\hbar \omega _{\text{s}},  \label{HBoson}
\end{eqnarray}%
where the use is made of the relations\cite{Bethe,Hulthen}, $\sigma _{\alpha
}^{-}\sigma _{\beta }^{+}=\hat{b}_{\beta }^{\dagger }\hat{b}_{\alpha }$, $%
\sigma _{\alpha }^{+}\sigma _{\beta }^{-}=\hat{b}_{\alpha }^{\dagger }\hat{b}%
_{\beta }$, $\hat{a}^{\dagger }\sigma _{\alpha }^{-}=\hat{a}^{\dagger }\hat{b%
}_{\alpha }$ and $\hat{a}\sigma _{\alpha }^{+}=\hat{b}_{\alpha }^{\dagger }%
\hat{a}$. The photon couples equally with each atom with the coupling
strength $\hbar g$ in (\ref{HBoson}), which implies that the additional
coupling due to the photon is highly nonlocal. Hence, it is a nontrivial
problem whether topological properties keep to hold in the presence of the
coupling with a photon.

\textbf{Cavity-coupled SSH model: } By setting $J_{\alpha \beta }=J_{A}$\ or 
$J_{B}$\ alternatively in a one-dimensional chain as in Fig.\ref%
{FigKagomeNanodiskIllust}(a), we obtain the dimerized XX model\cite%
{Taylor,Derz,Stolze},%
\begin{eqnarray}
H_{\text{spin-SSH}} &=&-\sum_{\alpha }[J_{A}\left( \sigma _{2\alpha
-1}^{-}\sigma _{2\alpha }^{+}+\sigma _{2\alpha -1}^{+}\sigma _{2\alpha
}^{-}\right)  \notag \\
&&+J_{B}\left( \sigma _{2\alpha }^{-}\sigma _{2\alpha +1}^{+}+\sigma
_{2\alpha }^{+}\sigma _{2\alpha +1}^{-}\right) ]  \notag \\
&&+\hbar \omega _{\text{s}}\left( \sum_{\alpha }\frac{\sigma _{\alpha
}^{+}\sigma _{\alpha }^{-}}{2}-1\right) ,
\end{eqnarray}%
from the interaction part of the Hamiltonian (\ref{HMag}). The corresponding
bosonic Hamiltonian together with the photon coupling and the photon loss
reads%
\begin{eqnarray}
H_{\text{cavity-SSH}} &=&\hbar \omega _{0}\hat{a}^{\dagger }\hat{a}%
-\sum_{\alpha }[J_{A}\left( \hat{b}_{2\alpha -1}^{\dagger }\hat{b}_{2\alpha
}+\hat{b}_{2\alpha }^{\dagger }\hat{b}_{2\alpha -1}\right)  \notag \\
&&+J_{B}\left( \hat{b}_{2\alpha }^{\dagger }\hat{b}_{2\alpha +1}+\hat{b}%
_{2\alpha +1}^{\dagger }\hat{b}_{2\alpha }\right) ]-\hbar \omega _{\text{s}}
\notag \\
&&+\hbar g\sum_{\alpha }\left( \hat{a}^{\dagger }\hat{b}_{\alpha }+\hat{b}%
_{\alpha }^{\dagger }\hat{a}\right) -\frac{i\hbar }{2}\gamma \hat{a}%
^{\dagger }\hat{a}.  \label{BasicSSH}
\end{eqnarray}%
This is the basic Hamiltonian we analyze. See Eq.(S4) in Supplemental
Material with respect to the photon loss term $-\frac{1}{2}i\hbar \gamma 
\hat{a}^{\dagger }\hat{a}$. We set $J_{A}=J(1+\lambda )$ and $%
J_{B}=J(1-\lambda )$ with $\lambda $ the dimerization.\ We take $\hbar
\gamma /J=0.2$\ in numerical simulations throughout the paper.

\begin{figure}[t]
\centerline{\includegraphics[width=0.48\textwidth]{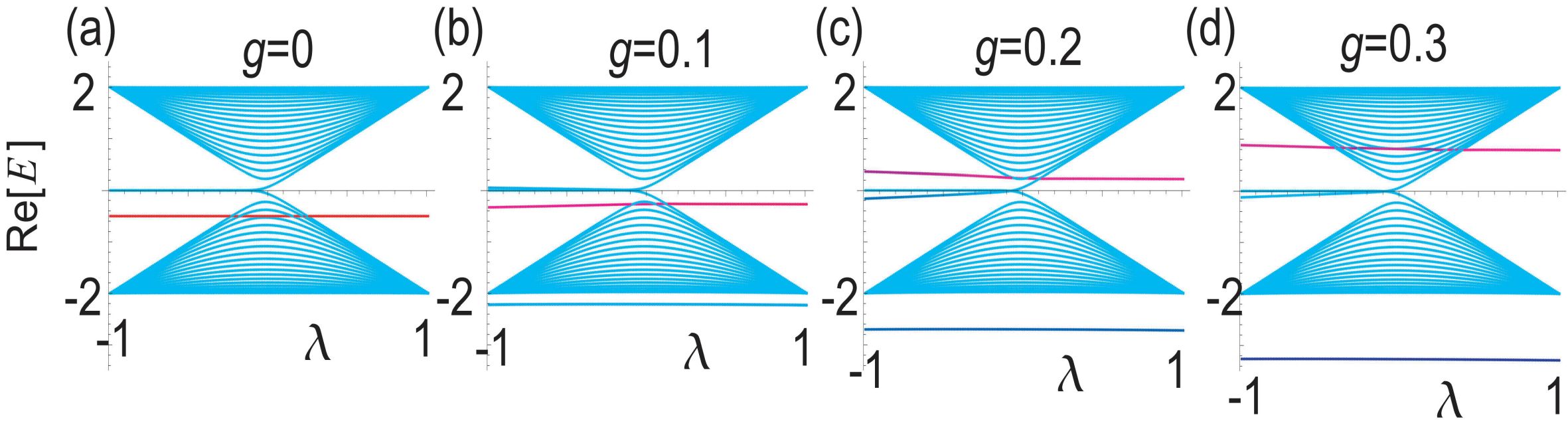}}
\caption{Energy spectrum as a function of $\protect\lambda $, describing
bulk bands in cyan, the topological edges in cyan and the photon in red. (a) 
$\hbar g/J=0$, (b) $\hbar g/J$, (c) $\hbar g/J=0.2$ and (d) $\hbar g/J=0.3$.
The vertical axis is the energy in units of $J$. We have set $N=80$. }
\label{FigSSHBand}
\end{figure}

In the absence of the photon coupling ($g=0$), the Hamiltonian matrix is
identical to the SSH\ model. The system is trivial for $\left\vert
J_{A}\right\vert >\left\vert J_{B}\right\vert $\ ($\lambda >0$) and
topological for $\left\vert J_{A}\right\vert <\left\vert J_{B}\right\vert $\
($\lambda <0$). We show the energy spectrum for $g=0$ as a function of $%
\lambda $\ in Fig.\ref{FigSSHBand}(a). In the trivial phase ($\lambda >0)$,
there is no zero-energy edge state. In the topological phase ($\lambda <0$),
there are two zero-energy edge states localized at the left and right edges.
They form the symmetric state and the antisymmetric state. The photon state
is present as indicated by a red line parallel to the $\lambda $-axis.

The energy spectrum is given for $g\neq 0$ as a function of $\lambda $\ in
Fig.\ref{FigSSHBand}(b)$\sim $(d). For $\lambda <0$, the symmetric
topological edge state slightly acquires a nonzero energy due to the photon
coupling, while the antisymmetric topological edge state remains precisely
at zero energy because the antisymmetric state does not couple with the
photon. Furthermore, for all $\lambda $, one flat state is detached from the
bulk band, which we call the detached state. The photon couples only with
the symmetric bulk state in the Hamiltonian (\ref{BasicSSH}). In addition,
the photon state is present as indicated by a red line tilted slightly
against the $\lambda $-axis.

\begin{figure}[t]
\centerline{\includegraphics[width=0.48\textwidth]{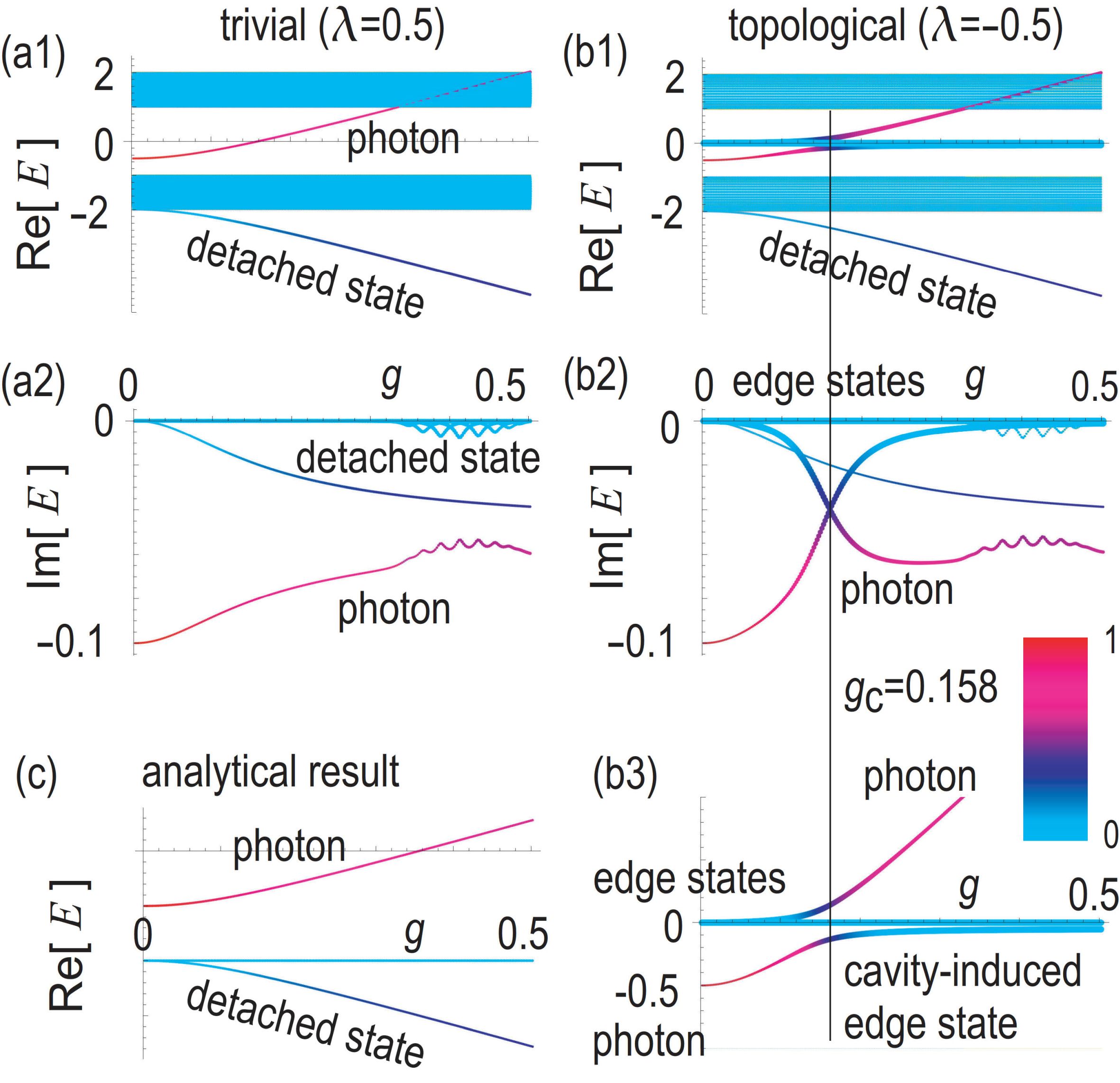}}
\caption{Energy spectrum as a function of the coupling strength $g$ in the
trivial phase with $\protect\lambda =0.5$ for (a1)$\sim $(a2) and in the
topological phase with $\protect\lambda =-0.5$ for (b1)$\sim $(b3). The
photon energy is taken so that $\hbar \left( \protect\omega _{0}-\protect%
\omega _{\text{s}}\right) =-0.5J$. Color palette shows the portion of the
photon amplitude $\left\langle \hat{a}^{\dagger }\hat{a}\right\rangle $,
where red and cyan indicate that it is $\left\langle \hat{a}^{\dagger }\hat{a%
}\right\rangle =1$ and $\left\langle \hat{a}^{\dagger }\hat{a}\right\rangle
=0$, respectively. (b3) A detailed band structure at the anticrossing point
in the topological phase, where there are three branches, the
positive-energy branch, the zero-energy branch and the negative-energy
branch. The vertical axis is the energy in units of $J$. \ We have set $N=80$%
. (c) Analytical result of Eq.(\protect\ref{EqAna}).}
\label{FigSSH}
\end{figure}

In order to make clear these properties, we show the energy spectrum as a
function of the coupling constant $g$ in Fig.\ref{FigSSH}. For definiteness,
we take the energy $\hbar \left( \omega _{0}-\omega _{\text{s}}\right) $ in
the gap between the two bulk bands as in Fig.\ref{FigSSH}(a1). We obtain
similar results for other cases of the photon energy, about which we show in
Supplemental Material: See Fig.(S1).

\begin{figure*}[t]
\centerline{\includegraphics[width=0.92\textwidth]{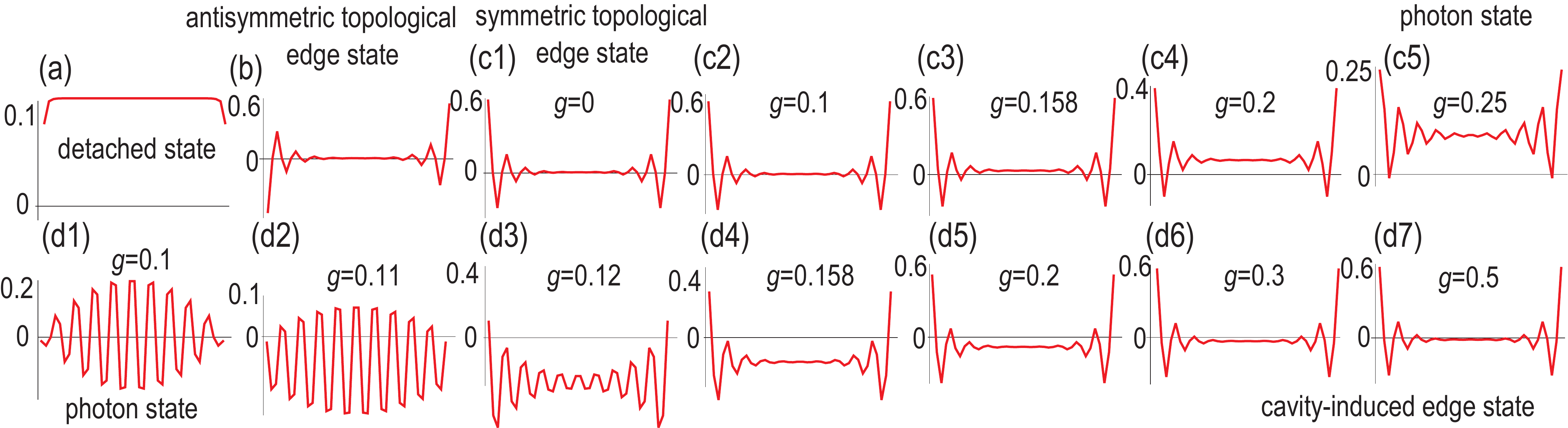}}
\caption{(a) Real part Re$\left[ \protect\psi _{\protect\alpha }\right] $ of
the eigenfunction of the detached state. (b) Those of the antisymmetric
topological edge states at zero energy irrespective of $g$. (c1)$\sim $(c5)
Those of the negative-energy branch for various $g$. (d1)$\sim $(d7) Those
of the positive-energy branch for various $g$. The horizontal axis is the
site index.\ We have considered the topological phase ($\protect\lambda %
=-0.5 $). The anticrossing point is $\hbar g_{c}/J=0.158$. We have set $N=80$%
. }
\label{FigSSHEdge}
\end{figure*}
\begin{figure*}[t]
\centerline{\includegraphics[width=0.98\textwidth]{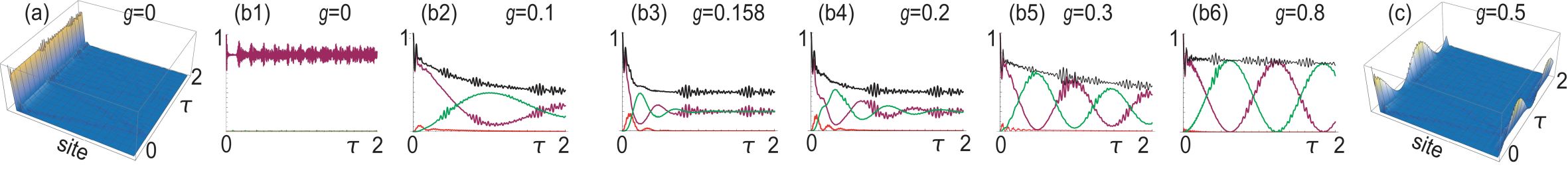}}
\caption{Quench dynamics of the boson number $\langle \hat{b}_{\protect%
\alpha }^{\dagger }\hat{b}_{\protect\alpha }\rangle $ starting from the left
edge state in the topological phase ($\protect\lambda =-0.5$) for the time
period $0<\protect\tau \equiv tJ/\hbar <2$, where (a) for $g=0$ and (c) for $%
g=0.5$. (b1)$\sim $(b6) Time evolution of the boson number $\langle \hat{b}%
_{1}^{\dagger }\hat{b}_{1}\rangle $ at the left edge and $\langle \hat{b}%
_{N}^{\dagger }\hat{b}_{N}\rangle $ at the right edge and the photon number $%
\left\langle \hat{a}^{\dagger }\hat{a}\right\rangle $ for the time period $%
0<tJ/\hbar <2$. Violet curves indicate the time evolution of the left-edge
site $\langle \hat{b}_{1}^{\dagger }\hat{b}_{1}\rangle $, green curves
indicate the time evolution of the right-edge site $\langle \hat{b}%
_{N}^{\dagger }\hat{b}_{N}\rangle $, red curves indicate the time evolution
of the photon state $\left\langle \hat{a}^{\dagger }\hat{a}\right\rangle $,
and black curves indicate the sum of the both-edge sites and the photon
state. Only the left edge state is excited for $g=0$, but both the edges are
excited for $g\neq 0$. Out-of-phase oscillations between the left and right
edges are suppressed as in (b3) around the anticrossing point $g_{\text{c}%
}=0.158$. We have set $N=80$. }
\label{FigSSHDynamics}
\end{figure*}

We first examine the trivial phase, where the spectrum consists four parts,
one photon state, two bulk bands and one detached state. See Fig.\ref{FigSSH}%
(a1) and (a2). See Fig.\ref{FigSSHEdge}(a) for the distribution of the real
part Re$\left[ \psi _{\alpha }\right] $ of the eigenfunction\ of the
detached state, showing that it is almost flat bulk state. The energy of the
photon state increases monotonously, while the energy of the detached state
decreases monotonously. We make an analytical study of the spectrum of the
detached state and the photon state later in the paragraph containing Eq.(%
\ref{EqAna}).

Intriguing phenomena emerge in the topological phase, where the photon
couples with the symmetric zero-energy edge state as in Fig.\ref{FigSSH}(b1)$%
\sim $(b3). A critical coupling $g_{\text{c}}$ emerges at which an
anticrossing of these two states occurs.

We identify the anticrossing point $g_{\text{c}}$\ in the real energy
spectrum [Fig.\ref{FigSSH}(b1)] from the crossing point in the imaginary
energy spectrum [Fig.\ref{FigSSH}(b2)]. A detailed structure of the
anticrossing is shown in Fig.\ref{FigSSH}(b3), where we observe three
branches, the positive-energy branch, the zero-energy branch and the
negative-energy branch.

First, we focus on the positive-energy branch, which starts from the
symmetric topological zero-energy edge state at $g=0$. As $g$\ increases,
its energy remains almost zero below the anticrossing point ($g<g_{\text{c}}$%
), but becomes positive for $g>g_{\text{c}}$, as in Fig.\ref{FigSSH}(b1) and
(b3). Correspondingly, the edge state is transformed into a bulk state as in
Fig.\ref{FigSSHEdge}(c1)$\sim $(c5). The symmetric topological zero-energy
edge state is transformed eventually into the photon state as $g$\ increases.

Second, the zero-energy blanch contains only the antisymmetric topological
edge state [Fig.\ref{FigSSHEdge}(b)], which is independent of the coupling
constant $g$.

Third, we focus on the negative-energy branch, which starts from a pure
photon state at $g=0$. As $g$\ increases, its energy increases toward zero
around the anticrossing point $g_{\text{c}}$, and becomes almost zero for $%
g>g_{\text{c}}$, as in Fig.\ref{FigSSH}(b1) and (b3). Correspondingly, the
bulk state is transformed into the symmetric edge state as in Fig.\ref%
{FigSSHEdge}(d1)$\sim $(d7).\ The photon state is transformed eventually
into the symmetric topological zero-energy edge state as $g$\ increases. It
is the cavity-induced edge state.

On the other hand, the structures of the two bulk bands is independent of
the coupling constant $g$. In particular, the bulk gap does not close, and
hence, the topological properties are robust. We verify this observation by
calculating the topological charge in Supplemental Material.

\textbf{Analytic study: } It is intriguing that one bulk state is detached
from the bulk band due to the interaction with a photon both in the trivial
and topological phases as in Fig.\ref{FigSSH}(a1) and (b1). In order to
understanding this phenomenon analytically, we make a study of a simple
model with $J_{\alpha \beta }=0$, where $N$\ zero-energy states couple with
one photon equally. The energy spectrum is analytically obtained as%
\begin{equation}
E/\hbar =\frac{\omega _{0}-i\gamma /2\pm \sqrt{\left( \omega _{0}-i\gamma
/2\right) ^{2}+4Ng^{2}}}{2},  \label{EqAna}
\end{equation}%
together with $N-1$ zero-energy level. It is plotted as a function of $g$ in
Fig.\ref{FigSSH}. These curves well explain the numerically obtained
cavity-induced energy spectrum in Fig.\ref{FigSSH}(a1) and (b1).

\textbf{Quench dynamics: } We study quench dynamics starting from the
left-edge site. We numerically solve the Schr\"{o}dinger equation $i\hbar 
\frac{d}{dt}\left\vert \psi \right\rangle =H_{\text{cavity-SSH}}\left\vert
\psi \right\rangle $ with Eq.(\ref{BasicSSH}) by imposing the initial
condition $\psi \left( t=0\right) =\delta _{\alpha ,1}$ on the amplitude.

We show the time evolution in Fig.\ref{FigSSHDynamics}. When there is no
coupling $g=0$ as in Fig.\ref{FigSSHDynamics}(a) and (b1), only the
amplitude at the left edge site is excited. As soon as $g\neq 0$, the
right-edge site is also excited via the photon coupling as in Fig.\ref%
{FigSSHDynamics}(b2)$\sim $(b6) and (c). The oscillations are out of phase
with equal strength between the left and right edges. However, it is notable
that the oscillations are suppressed around the anticrossing point $g_{\text{%
c}}$ as in Fig.\ref{FigSSHDynamics}(b3). This is because only the
antisymmetric edge state is at zero energy while the symmetric edge state
acquires a non-zero energy.

\textbf{Cavity-coupled Kagome second-order topological insulator: } We next
analyze the breathing Kagome XX model as illustrated in Fig.\ref%
{FigKagomeNanodiskIllust}(b), which is a generalization of the dimerized XX\
model to the breathing Kagome lattice. The Hamiltonian is given by 
\begin{eqnarray}
H_{\text{spin-Kagome}} &=&-\sum_{\left\{ A,B\right\} }[J_{A}\left( \sigma
_{A}^{-}\sigma _{B}^{+}+\sigma _{A}^{+}\sigma _{B}^{-}\right)  \notag \\
&&+J_{B}\left( \sigma _{B}^{-}\sigma _{A}^{+}+\sigma _{B}^{+}\sigma
_{A}^{-}\right) ]  \notag \\
&&+\hbar \omega _{\text{s}}\left( \sum_{\alpha }\frac{\sigma _{\alpha
}^{+}\sigma _{\alpha }^{-}}{2}-1\right) ,
\end{eqnarray}%
where the XX interaction exists in the nearest-neighbor sites in the
breathing Kagome lattice. The corresponding Hamiltonian is a second-order
topological insulator model\cite{EzawaKagome} with the photon coupling and
the photon loss,%
\begin{eqnarray}
H_{\text{cavity-Kagome}} &=&\hbar \omega _{0}\hat{a}^{\dagger }\hat{a}%
-\sum_{\left\{ A,B\right\} }[J_{A}\left( \hat{b}_{A}^{\dagger }\hat{b}_{B}+%
\hat{b}_{B}^{\dagger }\hat{b}_{A}\right)  \notag \\
&&+J_{B}\left( \hat{b}_{B}^{\dagger }\hat{b}_{A}+\hat{b}_{A}^{\dagger }\hat{b%
}_{B}\right) ]+\hbar \omega \hat{b}_{\alpha }^{\dagger }\hat{b}_{\alpha
}-\hbar \omega _{\text{s}}  \notag \\
&&+\hbar g\sum_{\alpha }\left( \hat{a}^{\dagger }\hat{b}_{\alpha }+\hat{b}%
_{\alpha }^{\dagger }\hat{a}\right) -\frac{i\hbar }{2}\gamma \hat{a}%
^{\dagger }\hat{a}.  \label{BasicKagome}
\end{eqnarray}%
This is the basic Hamiltonian.

\begin{figure}[t]
\centerline{\includegraphics[width=0.48\textwidth]{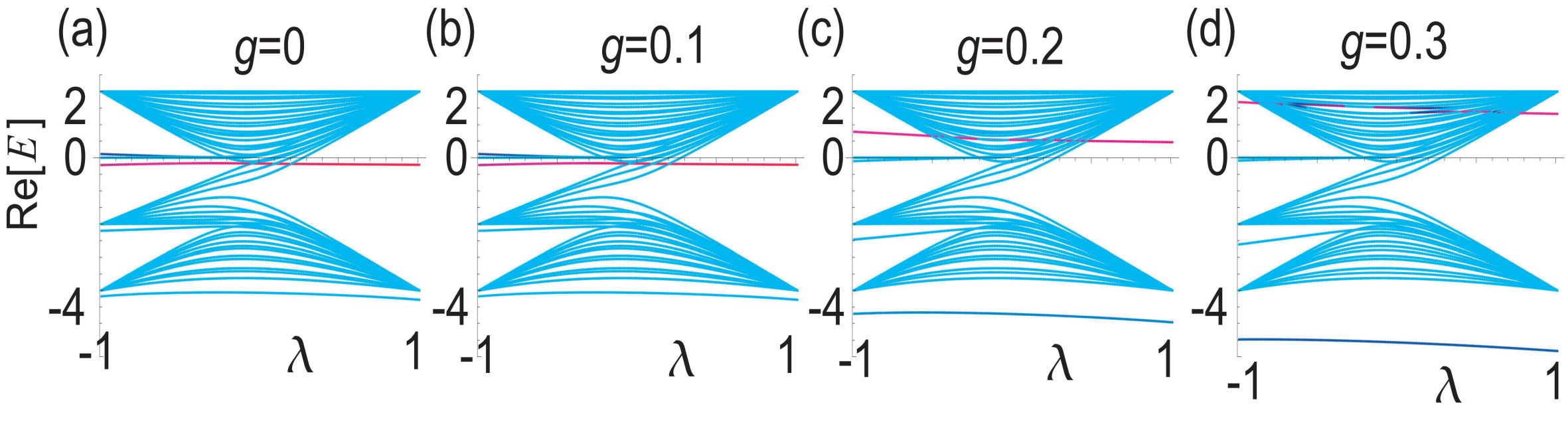}}
\caption{Energy spectrum as a function of $\protect\lambda $ in the
breathing Kagome model, describing three bulk bands in cyan, the topological
corners in cyan and the photon in red. (a) $\hbar g/J=0$, (b) $\hbar g/J=0.1$%
, (c) $\hbar g/J=0.2$ and (d) $\hbar g/J=0.3$. The vertical axis is the
energy in units of $J$. \ We have set $N=108$.}
\label{FigKagomeBand}
\end{figure}

\begin{figure}[t]
\centerline{\includegraphics[width=0.48\textwidth]{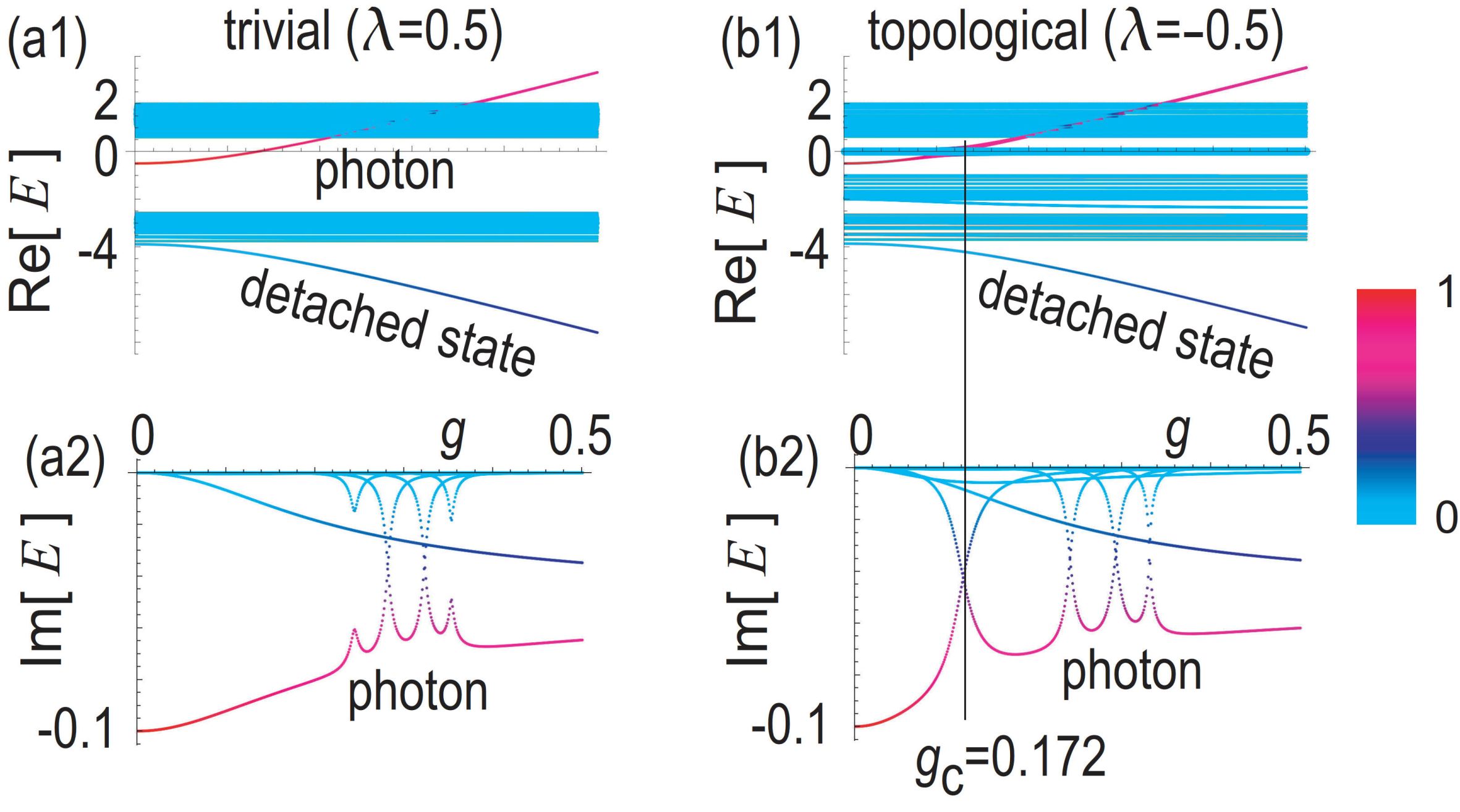}}
\caption{Energy spectrum of the breathing Kagome model as a function of the
coupling strength $g$ in the trivial phase with $\protect\lambda =0.5$ for
(a1)$\sim $(a2) and in the topological phase with $\protect\lambda =-0.5$
for (b1)$\sim $(b2). The photon energy is taken so that $\hbar \left( 
\protect\omega _{0}-\protect\omega _{\text{s}}\right) =0.5J$. Color palette
shows the portion of the photon amplitude $\left\langle \hat{a}^{\dagger }%
\hat{a}\right\rangle $, where red and cyan indicate that it is $\left\langle 
\hat{a}^{\dagger }\hat{a}\right\rangle =1$ and $\left\langle \hat{a}%
^{\dagger }\hat{a}\right\rangle =0$, respectively. A detailed band structure
at the anticrossing point in the topological phase is almost identical to
Fig.\protect\ref{FigSSH}(b3). Only difference is that the zero-energy blanch
contains two topological corner states except for the C$_{3}$ symmetric one.
The vertical axis is the energy in units of $J$. \ We have set $N=108$.}
\label{FigKagome}
\end{figure}

The analysis is similarly done as in the cavity-coupled SSH model. The
system is trivial for $J_{A}>J_{B}$\ ($\lambda >0$) and topological for $%
J_{A}<J_{B}$\ ($\lambda <0$). We show the energy spectrum for $g=0$ as a
function of $\lambda $\ in Fig.\ref{FigKagomeBand}(a). In the trivial phase (%
$\lambda >0)$, there is no zero-energy edge state. In the topological phase (%
$\lambda <0$), there are three zero-energy topological corner states
localized at the three corners. One of them forms the C$_{3}$ symmetric
state. In addition, the photon state is present as indicated by a red line
parallel to the $\lambda $-axis.

The energy spectrum is given for $g\neq 0$ as a function of $\lambda $\ in
Fig.\ref{FigKagomeBand}(b)$\sim $(d). For $\lambda <0$, the C$_{3}$
symmetric topological corner state slightly acquires a nonzero energy due to
the photon coupling, while the other two topological corner states remain at
zero energy. Furthermore, for all $\lambda $, one flat state is detached
from the bulk band, which is the C$_{3}$ symmetric bulk state.

We show the energy spectrum as a function of the coupling constant $g$ in
Fig.\ref{FigKagome}. For definiteness, we take the photon energy $\hbar
\left( \omega _{0}-\omega _{\text{s}}\right) $ in the gap between the two
bulk bands near the zero-energy level as in Fig.\ref{FigKagome}(a1) and
(b1). In the trivial phase the spectrum consists four parts, one photon
state, two bulk bands and one detached state: See Fig.\ref{FigKagome}(a1)
and (a2). The energy of the photon state increases monotonously while the
energy of the detached state decreases monotonously.

On the other hand, in the topological phase the photon state anticrosses the
C$_{3}$ symmetric corner state as in Fig.\ref{FigKagome}(b1). The
anticrossing point $g_{\text{c}}$\ in the real energy spectrum is identified
from the crossing point in the imaginary energy spectrum as in Fig.\ref%
{FigKagome}(b2). A detailed structure of the anticrossing is very similar to
the one in the SSH model shown in Fig.\ref{FigSSH}(b3). We observe three
branches, the positive-energy branch, the zero-energy branch and the
negative-energy branch. The main difference is that there are two
zero-energy corner states in the zero-energy branch. In the positive-energy
branch, the C$_{3}$ symmetric edge state is transformed eventually into the
photon state as $g$\ increases. In the negative-energy branch, the photon
state is transformed eventually into the C$_{3}$ symmetric topological
zero-energy edge state as $g$\ increases. It is the cavity-induced corner
state.

\textbf{Discussion: } The boson numbers $\langle \hat{b}_{\alpha }^{\dagger }%
\hat{b}_{\alpha }\rangle $ of superconducting qubits are experimentally
accessible by quantum non-demolition measurements\cite{Lupa}. On the other
hand, the photon number $\left\langle \hat{a}^{\dagger }\hat{a}\right\rangle 
$ is experimentally observable in superconducting circuits\cite{Schust}.
Another way is the use of magnets, where strong photon-magnon coupling is
realized\cite{Imag,Soykal,Ams}. There is a possibility that our system is
realized in magnets.

This work is supported by CREST, JST (Grants No. JPMJCR20T2) and
Grants-in-Aid for Scientific Research from MEXT KAKENHI (Grant No.
23H00171). 


\clearpage\newpage
\onecolumngrid
\def\theequation{S\arabic{equation}}
\def\thefigure{S\arabic{figure}}
\def\thesubsection{S\arabic{subsection}}
\setcounter{figure}{0}
\setcounter{equation}{0}
\setcounter{section}{0} 

\begin{center}
\textbf{\Large Supplemental Material} \bigskip \bigskip

\textbf{\large Cavity-induced topological edge and corner states }\bigskip

{Motohiko Ezawa}

{Department of Applied Physics, The University of Tokyo, 7-3-1 Hongo, Tokyo
113-8656, Japan}
\end{center}

\section{Photon loss}

Because the cavity is an open quantum system it is necessary to include the
effect of the photon loss . The Lindblad equation for the density matrix $%
\rho $\ reads%
\begin{equation}
\frac{d\rho }{dt}=-\frac{i}{\hbar }\left[ H,\rho \right] +\gamma \left(
L\rho L^{\dagger }-\frac{1}{2}\left\{ L^{\dagger }L,\rho \right\} \right) .
\end{equation}%
where $L$ is the Lindblad operator describing the dissipation $\gamma $.
This equation is rewritten in the form of%
\begin{equation}
\frac{d\rho }{dt}=-\frac{i}{\hbar }\left( H_{\text{eff}}\rho -\rho H_{\text{%
eff}}^{\dagger }\right) +\gamma L\rho L^{\dagger },
\end{equation}%
where $H_{\text{eff}}$ is a non-Hermitian effective Hamiltonian defined by%
\begin{equation}
H_{\text{eff}}\equiv H-\frac{i\hbar \gamma }{2}L^{\dagger }L.
\end{equation}%
The photon loss is described by the Lindblad operator as $L=\hat{a}$. The
non-Hermitian effective Hamiltonian reads%
\begin{equation}
H_{\text{eff}}\equiv H-\frac{i\hbar }{2}\gamma \hat{a}^{\dagger }\hat{a}.
\label{PhotonLoss}
\end{equation}%
The imaginary part of the energy spectrum is proportional to the photon
amplitude $\left\langle \hat{a}^{\dagger }\hat{a}\right\rangle $, where the
dissipation occurs only in the photon state in the present model.

\section{Topological number}

The Hamiltonian in the momentum space reads%
\begin{equation}
\hat{H}\left( k\right) =\left( 
\begin{array}{ccc}
\hat{b}_{A}^{\dagger } & \hat{b}_{B}^{\dagger } & \hat{a}^{\dagger }%
\end{array}%
\right) \left( 
\begin{array}{ccc}
0 & -\left( J_{A}+J_{B}e^{-iak}\right) & \hbar g\delta \left( k\right) \\ 
-\left( J_{A}+J_{B}e^{-iak}\right) & 0 & \hbar g\delta \left( k\right) \\ 
\hbar g\delta \left( k\right) & \hbar g\delta \left( k\right) & \hbar \omega
_{0}%
\end{array}%
\right) \left( 
\begin{array}{c}
\hat{b}_{A} \\ 
\hat{b}_{B} \\ 
\hat{a}%
\end{array}%
\right) .
\end{equation}%
The winding number for the boson $\hat{b}_{\alpha }$ is given by%
\begin{equation}
W\equiv \frac{1}{2\pi ai}\int_{0}^{2\pi }\left\langle \psi _{b}\right\vert 
\frac{d}{dk}\left\vert \psi _{b}\right\rangle dk.
\end{equation}%
where $\left\vert \psi _{b}\right\rangle $ is the right eigenfunction and $%
\left\langle \psi _{b}\right\vert $ is the left eigenfunction of the boson $%
\hat{b}_{\alpha }$. We separate it as%
\begin{eqnarray}
W &\equiv &W_{1}+W_{2}+W_{3}, \\
W_{1} &\equiv &\frac{1}{2\pi ai}\lim_{\varepsilon \rightarrow
0}\int_{0+\varepsilon }^{2\pi -\varepsilon }\left\langle \psi
_{b}\right\vert \frac{d}{dk}\left\vert \psi _{b}\right\rangle dk, \\
W_{2} &\equiv &\frac{1}{2\pi ai}\lim_{\varepsilon \rightarrow
0}\int_{0}^{\varepsilon }\left\langle \psi _{b}\right\vert \frac{d}{dk}%
\left\vert \psi _{b}\right\rangle dk, \\
W_{3} &\equiv &\frac{1}{2\pi ai}\lim_{\varepsilon \rightarrow 0}\int_{2\pi
-\varepsilon }^{2\pi }\left\langle \psi _{b}\right\vert \frac{d}{dk}%
\left\vert \psi _{b}\right\rangle dk.
\end{eqnarray}%
$W_{1}$ is calculated by using the eigenfunction of the two-band Hamiltonian%
\begin{equation}
H\left( k\right) =-\left( 
\begin{array}{cc}
0 & J_{A}+J_{B}e^{-iak} \\ 
J_{A}+J_{B}e^{-iak} & 0%
\end{array}%
\right) ,  \label{HTwo}
\end{equation}%
which is identical to the original SSH model. Eigenenergies are%
\begin{equation}
E\left( k\right) =\pm \sqrt{J_{A}^{2}+J_{B}^{2}+2J_{A}J_{B}\cos ak},
\end{equation}%
and the eigenfunctions are%
\begin{equation}
\psi \left( k\right) =\frac{1}{\sqrt{2}}\left( 1,\mp \frac{J_{A}+J_{B}e^{iak}%
}{\left\vert E\left( k\right) \right\vert }\right) .
\end{equation}%
Hence, we have $W_{1}=1$ for the topological phase ($\left\vert
J_{A}\right\vert <\left\vert J_{B}\right\vert $) and $W_{1}=0$ for the
trivial phase ($\left\vert J_{A}\right\vert >\left\vert J_{B}\right\vert $).

$W_{2}$ is rewritten as%
\begin{equation}
W_{2}\equiv \frac{1}{2\pi ai}\lim_{\varepsilon \rightarrow 0}\varepsilon
\left\langle \psi _{b}\left\vert \psi _{b}\right\rangle \right. dk,
\end{equation}%
where $\left\vert \psi _{b}\right\rangle $ is the eigenfunction of (\ref%
{HTwo}) and $\left\langle \psi _{b}\right\vert $ is the eigen function of%
\begin{equation}
H\left( 0\right) =\left( 
\begin{array}{ccc}
0 & -\left( J_{A}+J_{B}\right) & \hbar g \\ 
-\left( J_{A}+J_{B}\right) & 0 & \hbar g \\ 
\hbar g & \hbar g & \hbar \omega _{0}%
\end{array}%
\right) ,
\end{equation}%
where the eigenenergies are%
\begin{equation}
E\left( 0\right) =J_{A}+J_{B},\frac{-\left( J_{A}+J_{B}\right) +\hbar \omega
^{\prime }\sqrt{\left( J_{A}+J_{B}+\hbar \omega _{0}\right) ^{2}+8\hbar
^{2}g^{2}}}{2},
\end{equation}%
and the corresponding eigenfunctions are%
\begin{equation}
\psi \left( 0\right) =\frac{1}{\sqrt{2}}\left( -1,1,0\right) ,\frac{1}{\sqrt{%
1+2\left\vert f_{\pm }\right\vert ^{2}}}\left( f_{\pm },f_{\pm },1\right) ,
\end{equation}%
with%
\begin{equation}
f_{\pm }\equiv \frac{J_{A}+J_{B}-\hbar \omega _{0}\pm \sqrt{\left(
J_{A}+J_{B}-\hbar \omega _{0}\right) ^{2}+8\hbar ^{2}g^{2}}}{4\hbar g}.
\end{equation}%
We take the part of boson $\hat{b}$, which is given by%
\begin{equation}
\psi \left( 0\right) =\frac{1}{\sqrt{2}}\left( 1,1\right) .
\end{equation}%
We have the finite inner product 
\begin{equation}
\lim_{\varepsilon \rightarrow +0}\left\langle \psi _{b}\left( \varepsilon
\right) \left\vert \psi _{b}\left( 0\right) \right\rangle \right. =1,
\end{equation}%
and hence, we obtain $W_{2}=0$. In the similar way, we have $W_{3}=0$. As a
result, we conclude $W=1$ for the topological phase ($\left\vert
J_{A}\right\vert <\left\vert J_{B}\right\vert $) and $W=0$ for the trivial
phase ($\left\vert J_{A}\right\vert >\left\vert J_{B}\right\vert $). Hence,
there is no topological phase transition induced by the photon coupling.

\begin{figure}[t]
\centerline{\includegraphics[width=0.68\textwidth]{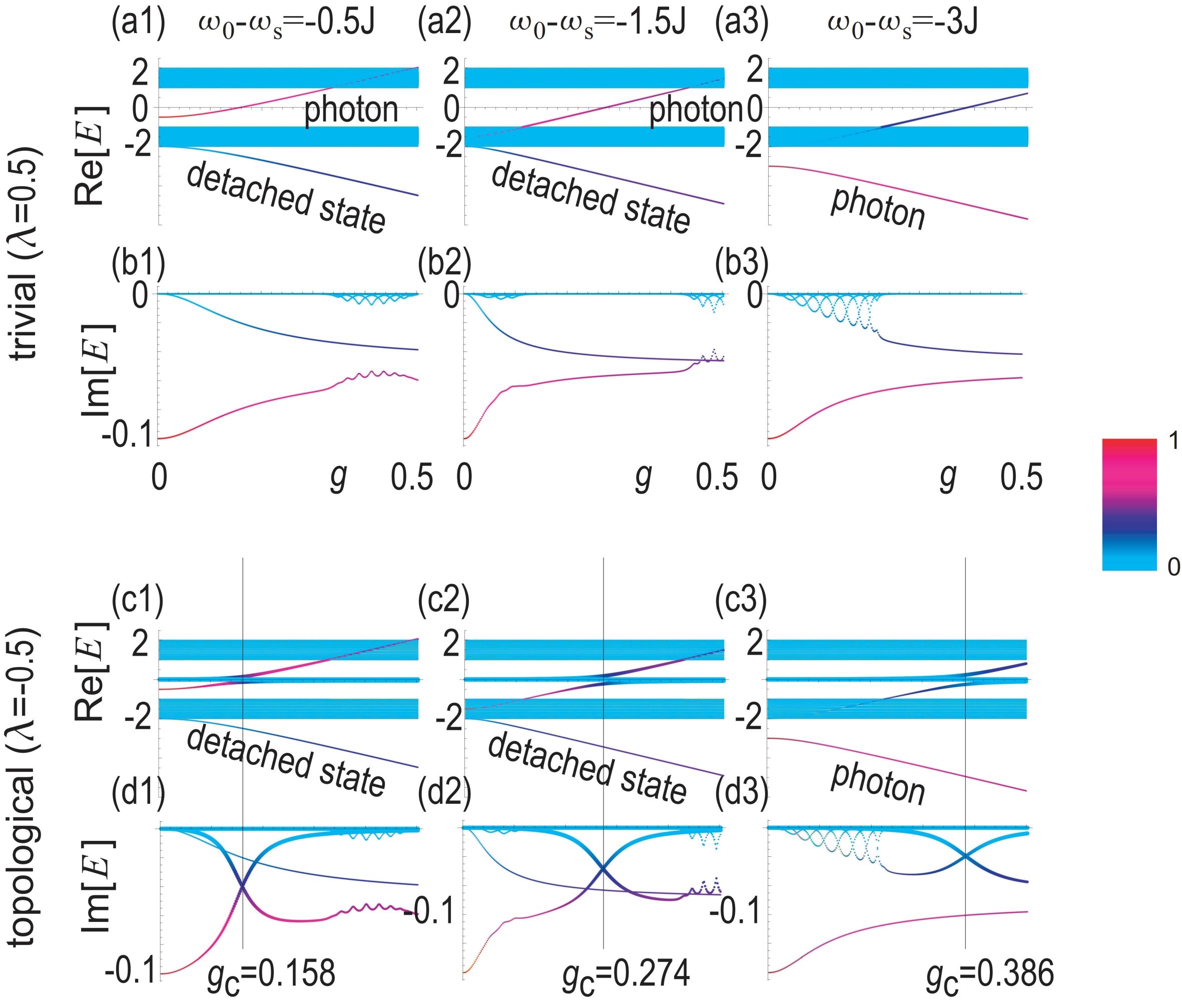}}
\caption{Energy spectrum as a function of the coupling strength $g$ in the
trivial phase with $\protect\lambda =0.5$ for (a1)$\sim $(b3) and in the
topological phase with $\protect\lambda =-0.5$ for (c1)$\sim $(d3). The
photon energy is taken as (*1) $\hbar \left( \protect\omega _{0}-\protect%
\omega _{\text{s}}\right) =-0.5J$, (*2) $\hbar \left( \protect\omega _{0}-%
\protect\omega _{\text{s}}\right) =-1.5J$ and (*3) $\hbar \left( \protect%
\omega _{0}-\protect\omega _{\text{s}}\right) =-3J$. Color palette shows the
portion of the photon amplitude $\left\langle \hat{a}^{\dagger }\hat{a}%
\right\rangle $, where red and cyan indicate that it is $\left\langle \hat{a}%
^{\dagger }\hat{a}\right\rangle =1$ and $\left\langle \hat{a}^{\dagger }\hat{%
a}\right\rangle =0$, respectively. A detailed band structure at the
anticrossing point is shown in Fig.\protect\ref{FigSSH}(b3). The vertical
axis is the energy in units of $J$. \ We have set $N=108$.}
\label{FigSSH0}
\end{figure}

\section{Cavity-coupled SSH model}

We investigate the cavity-coupled SSH model. There are three cases with
respect to the energy of the photon. (i) It is taken in the gap between the
two bulk bands as in Fig.\ref{FigSSH0}(a1). (ii) It is taken within the
lower bulk band as in Fig.\ref{FigSSH0}(a2). (iii) It is taken smaller than
the lower limit of the lower bulk band as in Fig.\ref{FigSSH0}(a3). We have
analyzed the case (i) in the main text. By comparing Fig.\ref{FigSSH0}(a1)
and (a2), the case (ii) is essentially the same as the case (i).

There are some different features in the case (iii). However, when we
interchange the roles of the photon state and the detached state, all
results follows as they are. In the trivial phase, for instance, the energy
of the detached state decreases monotonously while the energy of the photon
state increases monotoneously as in Fig.\ref{FigSSH0}(c3). In the
topological phase, the energies of the detached state and the topological
zero-energy edge state anticross at $g_{\text{c}}$. Furthermore, the
detached state is transformed into the symmetric topological zero-energy
edge state as $g$\ increases, and vice versa as in Fig.\ref{FigSSH0}(c3).

\end{document}